\documentclass[fleqn,1p,preprint]{elsarticle}

\makeatletter
\def\ps@pprintTitle{%
 \let\@oddhead\@empty
 \let\@evenhead\@empty
 \def\@oddfoot{\centerline{\thepage}}%
 \let\@evenfoot\@oddfoot}
\makeatother

\usepackage{amsmath,graphicx}

\usepackage[english]{babel}
\usepackage[latin1]{inputenc}
\usepackage[T1]{fontenc}
\usepackage{amsmath,graphicx}
\usepackage{amssymb,amsmath,epsfig,url,mathrsfs} 
\usepackage{algorithm,algorithmic}
\usepackage[usenames, dvipsnames]{color}
\usepackage[framemethod=TikZ]{mdframed}
\usepackage{xcolor}
\usepackage{listings}
\usepackage{subcaption}

\usepackage{ragged2e}
\usepackage{array}

\mdfdefinestyle{MyFrame}{%
    linecolor=black,
    outerlinewidth=.5pt,
    roundcorner=5pt,
    innertopmargin=\baselineskip,
    innerbottommargin=\baselineskip,
    innerrightmargin=5pt,
    innerleftmargin=0pt,
    backgroundcolor=gray!10!white}
\definecolor{codegreen}{rgb}{0,0.6,0}
\definecolor{codegray}{rgb}{0.5,0.5,0.5}
\definecolor{codepurple}{rgb}{0.58,0,0.82}
\definecolor{backcolour}{rgb}{0.95,0.95,0.92}

\lstdefinestyle{mystyle}{
    backgroundcolor=\color{backcolour},
    commentstyle=\color{codegreen},
    keywordstyle=\color{blue},
    numberstyle=\tiny\color{codegray},
    stringstyle=\color{codepurple},
    basicstyle=\footnotesize,
    breakatwhitespace=false,
    breaklines=true,
    captionpos=b,
    keepspaces=true,
    numbers=none,
    numbersep=5pt,
    showspaces=false,
    showstringspaces=false,
    showtabs=false,
    tabsize=2
}

\usepackage{amsmath}
\usepackage{graphicx}
\usepackage[colorinlistoftodos]{todonotes}
\usepackage[colorlinks=true, allcolors=blue]{hyperref}

\DeclareMathAlphabet{\mathpzc}{OT1}{pzc}{m}{it}

\newcommand{\revcolor}{black}

\journal{Computer Speech \& Language}

\begin{document}

\begin{frontmatter}

\title{Voice Mimicry Attacks Assisted by\\Automatic Speaker Verification \footnote{\copyright 2019. This manuscript version is made available under the CC-BY-NC-ND 4.0 license:\\\url{http://creativecommons.org/licenses/by-nc-nd/4.0/}}}

\author[uef]{Ville Vestman\corref{cor2}}
\ead{vvestman@cs.uef.fi}

\author[uef]{Tomi Kinnunen\corref{cor1}}
\ead{tkinnu@cs.joensuu.fi}

\author[uef]{Rosa Gonz\'alez Hautam\"aki}
\ead{rgonza@cs.uef.fi}

\author[inria]{Md Sahidullah}
\ead{md.sahidullah@inria.fr}

\address[uef]{School of Computing, University of Eastern Finland, FI-80101, Joensuu, Finland}

\address[inria]{Universit\'{e} de Lorraine, CNRS, Inria, LORIA, F-54000 Nancy, France}

\cortext[cor2]{\textcolor{\revcolor}{A part of the work of the first author was carried out during an intership at NEC.}}
\cortext[cor1]{Corresponding Author}

\begin{abstract}

\textcolor{\revcolor}{
In this work, we simulate a scenario, where a publicly available ASV system is used to enhance mimicry attacks against another closed source ASV system. In specific, ASV technology is used to perform a similarity search between the voices of recruited attackers (6) and potential target speakers (7,365) from VoxCeleb corpora to find the closest targets for each of the attackers. In addition, we consider `median', `furthest', and 'common' targets to serve as a reference points.}

\textcolor{\revcolor}{
Our goal is to gain insights how well similarity rankings transfer from the attacker's ASV system to the attacked ASV system, whether the attackers are able to improve their attacks by mimicking, and how the properties of the voices of attackers change due to mimicking. We address these questions through ASV experiments, listening tests, and prosodic and formant analyses. For the ASV experiments, we use i-vector technology in the attacker side, and x-vectors in the attacked side. For the listening tests, we recruit listeners through crowdsourcing.}

\textcolor{\revcolor}{
The results of the ASV experiments indicate that the speaker similarity scores transfer well from one ASV system to another. Both the ASV experiments and the listening tests reveal that the mimicry attempts do not, in general, help in bringing attacker's scores closer to the target's. A detailed analysis shows that mimicking does not improve attacks, when the natural voices of attackers and targets are similar to each other. The analysis of prosody and formants suggests that the attackers were able to considerably change their speaking rates when mimicking, but the changes in F0 and formants were modest. Overall, the results suggest that untrained impersonators do not pose a high threat towards ASV systems, but the use of ASV systems to attack other ASV systems is a potential threat.}

\end{abstract}


\end{frontmatter}


\section{Introduction}
\label{sec:intro}

\textcolor{\revcolor}{Security is of key importance in today's society where information processing gets increasingly digital, automated and lacks human-to-human communication}. We need \textcolor{\revcolor}{new ways to} protect our data records from unauthorized access. Alongside with the traditional means of user authentication, biometric technology has emerged as one of the potential solutions. The use of human voice for strong user authentication is attractive especially under remote, unattended scenarios and due to the readily available infrastructure (namely, telephones) to \textcolor{\revcolor}{scale it up easily}. 

Similar to the traditional means of user authentication, however, biometric systems are prone to malicious attacks by hackers. \textcolor{\revcolor}{It is no longer news, neither to the research community nor to the general public, that biometric systems} can be fooled through various \emph{representation attacks} ~\citep{Ratha2001,isopad}, also known as \emph{spoofing attacks}. A spoofing attack involves an adversary (attacker) who aims at masquerading oneself as another targeted user to gain illegitimate access to the targeted person's data. Unprotected \emph{automatic speaker verification} (ASV) systems \textcolor{\revcolor}{can be easily spoofed} using replay, voice conversion (VC) and text-to-speech (TTS) attacks  \citep{Wu2015-spoofing-survey}. Since the attacks are typically not perfect but contain either processing artifacts or display degraded audio quality, they can be detected to a certain extent. To this end, community-driven challenges such as ASVspoof \citep{Wu2015-asvspoof} and AVspoof \citep{Ergunay2015-realistic-voice-spoofing} were launched for an organized study of \emph{spoofing countermeasures}. \textcolor{\revcolor}{In the context of security, the continuous arms race between attacks and their defenses is well known} \citep{Biggio18-wild}: so as to develop effective countermeasures, it is necessary to understand the attacks. The speech synthesis community has independently launched \emph{voice conversion challenge\textcolor{\revcolor}{s}} \citep{Toda2016-VCC-challenge,Lorenzo2018-VCC18} to advance VC methods (though targeted primarily for human listeners rather than for ASV spoofing). \textcolor{\revcolor}{To sum up, within the past few years, active and dynamic communities both at the `attack' and `defense' sides of ASV have emerged}. There is now a far better understanding of the technology-based attacks and their defenses against ASV systems than half a decade ago --- see~\citep{SpringerHandbook} for an up-to-date review. 

\begin{figure}[!t]
\centering
\includegraphics[width=.90\columnwidth]{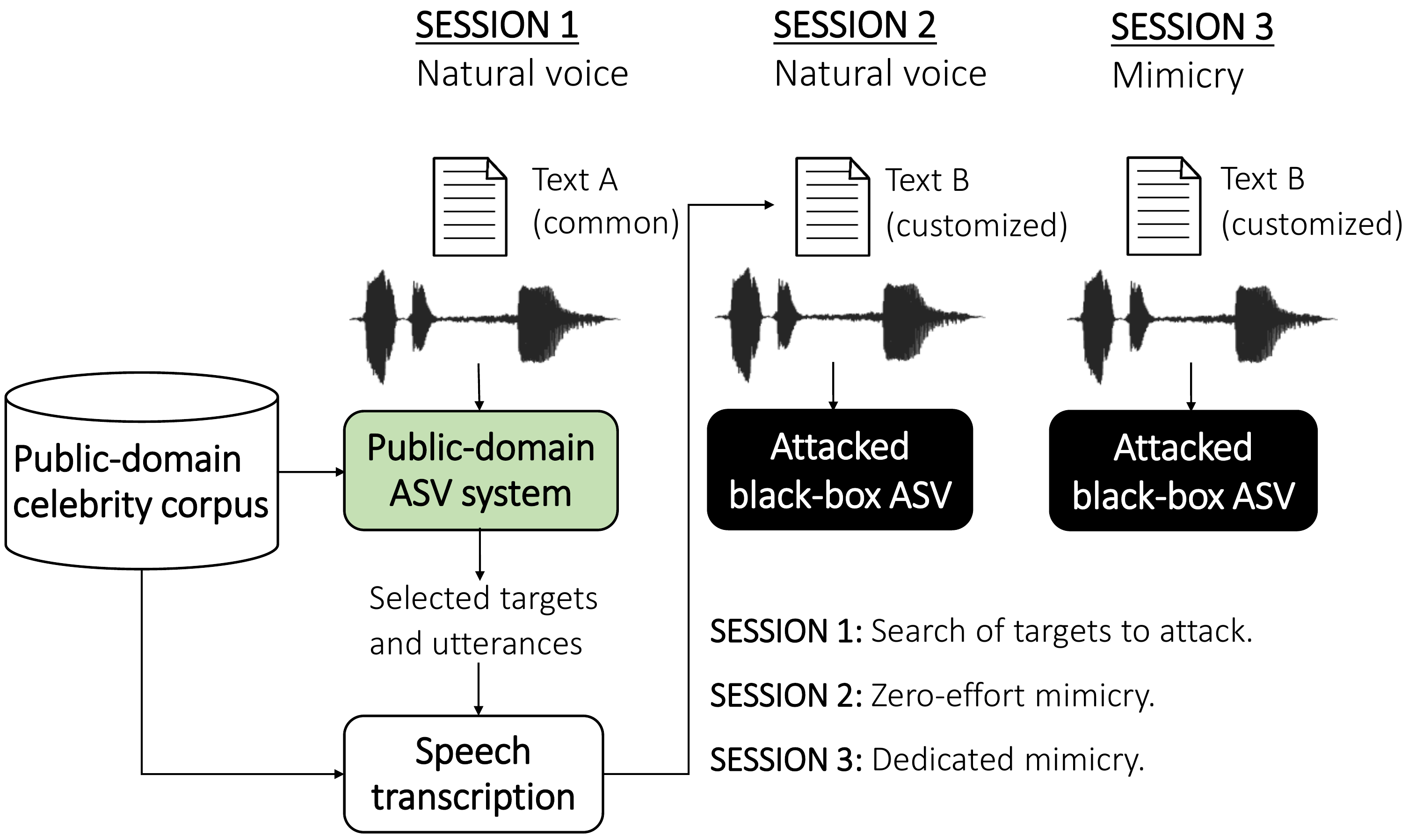}
\vspace{-0.1cm}
\caption{Automatic speaker verification (ASV) assisted mimicry attack: attacker uses a public-domain ASV system to select target speakers matched with his/her voice from a public celebrity data\textcolor{\revcolor}{base}. The attacker then practices target speaker mimicry, intended to attack another independently developed ASV system.}
\label{fig:study-overview}
\vspace{-0.3cm}
\end{figure}

In this study we focus on a nearly-forgotten ASV attack -- \emph{mimicry} (impersonation). Unlike the technology-induced attacks, mimicry involves \emph{human}-based modification of one's voice production. The question of recognizer vulnerability against mimicry was addressed at least around half a century ago \citep{Luck1969-cepstral,voicespectrograms1971} and has remained a cursory topic within the ASV field \citep{Lau-Vulnerability2004,Lau2005-mimicry,mariethoz2005-professional,eriksson2010disguised,Gonzalez2015-mimicry,Farrus2018}. While ASV vulnerability caused by technical attacks is widely reported, less (reliable) information is available on effectivess of mimicry, primarily due to adoption of small and proprietary datasets. The only conclusions that one can possibly extrapolate from the prior studies on mimicry effect against ASV is that the results depend on a specific study. This suggests that mimicry is less consistent attack compared to replay, VC and TTS that are repeatable reported to be successful in spoofing ASV systems.

The authors are aware of the difficulties in collecting mimicry data from professional artists \citep{Gonzalez2015-mimicry}, whose \textcolor{\revcolor}{prevalence} in the general population is arguably very low. Nonetheless, \emph{if} mimicry attacks could be shown to be a threat to ASV, it would be conceivably challenging to devise countermeasures: natural human speech lacks  processing artifacts that enable detection of technical attacks. Thus, we argue that it is important to keep mimicry also in the list of potential attacks against ASV. Besides the security aspect, mimicry could potentially help us in the design of better ASV methods for voice comparison. 

Of particular interest in this work are mimicry attacks against persons whose voice data is exposed in a public domain in large quantities --- such as celebrities or anyone streaming or uploading massive amounts of his/her videos to the Internet. In line with the recent EU's \emph{General Data Protection Regulation} (GDPR) \citep{GDPR}, intended to protect the privacy of individuals, it is important to assess potential risks associated with multimedia data in the public domain; we elaborate on this emerging problem further in Section \ref{sec:crowd-attacks}. Differently from most prior studies, we focus on \emph{technology-assisted} mimicry attacks. In specific, we use the ASV technology itself to identify potential target speakers to be subjected to mimicry attacks. The idea is to identify targets whose voice is \emph{a priori} similar to that of the attacker's voice in terms of acoustic parameters. The assumption is that nearby target speakers might be easier to mimic due to potentially fewer articulatory or voice source modifications required. Two related prior studies are \citep{Lau-Vulnerability2004} and \citep{Panjwani2014-crowd} which involve search of either targets \citep{Lau-Vulnerability2004} or attackers \citep{Panjwani2014-crowd} from a pool of candidates. The authors of \citep{Lau-Vulnerability2004} used a Gaussian mixture model (GMM) system to find closest, intermediate and furthest target speakers from YOHO corpus for two naive impersonators, leading to substantially increased false acceptance rate for the closest targets. In \citep{Panjwani2014-crowd}, the authors selected impersonators (rather than targets) through a commercial crowd-sourcing platform based on self-judgment and further refinement using ASV.

Our study can be seen as an attempt to reproduce the findings of \citep{Lau-Vulnerability2004} using up-to-date ASV technology and a far larger target candidate set ($7,365$ celebrities pooled from VoxCeleb1 \citep{Nagrani2017} and VoxCeleb2 \citep{Chung2018}). Besides the order of magnitude larger target speaker pool and adoption of state-of-the-art ASV systems, there is a key difference in the research methodology as well: unlike \citep{Lau-Vulnerability2004} that used a \emph{single} GMM recognizer, we include two \emph{different} ASV systems as illustrated in Fig. \ref{fig:study-overview}. We argue that it is unrealistic for the attacker to interact many times with the targeted ASV, as done in that past work. \textcolor{\revcolor}{In our attack model, therefore, the attacker uses an offline, publicly available \emph{substitute} ASV system to first identify which speakers to attack; ideally, the substitute system would behave  similar to the attacked ASV system.} This idea bears some resemblance to \emph{black box attacks} \citep{Papernot2017-practical-black-box} in \textcolor{\revcolor}{adversarial} machine learning \citep{Biggio18-wild}, though our adversary is not a machine learning algorithm but a human. Further, those methods use either classifier output score or decision to optimize the attacks, while we assume that the attacker receives no feedback from the attacked system in any form. Thus, we expect that our attacks are not strong, but we argue that they are \emph{realistic} given the abundance of both voice data and ASV implementations in the public domain. We seek to answer the question whether the use of ASV technology itself could increase the risk of an attacker being falsely accepted by (another) ASV system.

A preliminary version of this work appears in \cite{Kinnunen2018-can-we}. \textcolor{\revcolor}{Our preliminary findings in that work suggested a \emph{negative} result --- \emph{i.e.} that mimicry attempts, even when the target speakers were selected with automatic speaker identification, would not have left the attacked ASV systems vulnerable. We are not entirely content with just this finding, however --- we are interested to understand the reasons. To this end, the present work substantially extends \cite{Kinnunen2018-can-we} by  contrastive automatic, perceptual, prosody, and formant analyses. In particular, we include (i) \textbf{analysis of domain mismatch in ASV score domain} (presented in Section \ref{sec:asv_results}), (ii) a \textbf{human benchmark} of speaker similarity (presented in Section \ref{sec:perceptual-evaluation}), and (iii) \textbf{prosody and formant} analysis (presented in Section \ref{sec:prosody-analysis}). Additionally, (iv) Section \ref{sec:crowd-attacks} provides a broad background context to our work. None of the above were provided in \cite{Kinnunen2018-can-we}. The score domain analysis seeks to answer whether the negative finding might have been due to condition differences across our attacker and celebrity corpora. The human benchmark, implemented via crowdsourcing, serves for a reference point to the automatic methods}. Finally, the prosody \textcolor{\revcolor}{and formant} analyses serve to study changes in the speaking rate, fundamental frequency (F0), \textcolor{\revcolor}{and formants} induced by mimicry. Our hypothesis is that some of these `broad' speech parameters might be among the prominent cues that a naive mimic attempts to primarily modify towards the target speaker. \textcolor{\revcolor}{While this article is intended to be as self-contained as possible, the interested reader may consult additional online material \cite{CSL2018_additional_material} for further details about our text prompts and target speakers.}

\section{Attacks on speaker verification systems with found data}\label{sec:crowd-attacks}

The amount of personal data that people upload to the Internet increases year by year. Enabled by popular social media platforms and other picture/video sharing services, people upload (or stream) their self-portraits (selfies), voice samples and video clips much more easily --- perhaps more carelessly --- than in the past. The general public may be unaware that their face photos, videos and voice samples contain biometric traits and form potentially their `unique' identifiers\footnote{The authors argue that `unique' is a misleading term in the context of biometrics where decisions are not based on exact pattern matching but probabilistic reasoning.}. Somewhat paradoxically, of a specific concern is the rapidly advancing biometric technology \emph{itself}. The aim of biometric technology, similar to the traditional ways of user authentication, is to regulate access to a restricted domain. The basic premise is that a biometric database administrator (such as \textcolor{\revcolor}{the} police, \textcolor{\revcolor}{a} border control officer, or a bank) has sufficient security countermeasures to protect their biometric database and systems from being \textcolor{\revcolor}{hacked or tampered}. But what if the user decides to voluntarily expose his or her biometric data to the public? Very few of us would purposefully upload our credit card number or a photo-copy of our passport to a public website, but uploading our face and voice data \textcolor{\revcolor}{does not seem to concern many}. It is important to address the potential risk scenarios of misuse of personal data, and to make the general public aware of the potential risks \textcolor{\revcolor}{of} uploading their data to a public domain. Awareness on the potential risks among the professional community has increased due to initiatives such as EU's IC1206 COST action\footnote{\url{https://www.cost.eu/actions/IC1206}} that focused on de-identification and privacy protection of multimedia data (see \citep{Ribaric2016-deidentification-review} for a review). The overall picture is not yet complete, however, and human voice has received far less attention than image-based biometric traits in this context.

One potential risk is that biometric data that is not searchable or indexable using today's technology might become so tomorrow. Imagine a search engine that uses face or speaker recognition to cross-link someone's sensitive personal multimedia data --- such as sexually explicit photographs shared confidently with one's partner but leaked to a porn website; or a video portraying someone under the influence of drugs --- with his or her personal website or social media profile. Other risks could include fabricating a `digital clone' of someone using machine learning --- recent warning examples are provided by the so-called \emph{deepfakes}~\cite{chung2017you,liu2017unsupervised,suwajanakorn2017synthesizing}, realistic-appearing but fabricated or tampered videos portraying a targeted person created with the aid of deep learning (the interested reader is pointed to to~\cite{Deepfakes2018-looming-challenge} for a detailed review of potential societal, ethical and legal implications of deepfakes). In the context of speaker verification in specific, \cite{Lorenzo-Trueba2018-can-we-steal} addressed voice cloning of a well-known celebrity (the former US president Barack Obama). Even if the result was essentially negative (the cloned voice samples were detectable as artificial ones using a spoofing countermeasure), machine learning, including voice cloning techniques, do not stand still. 

As current machine learning models require large training sets, one may argue that persons who have more (and of technically higher-quality) data in the Internet might become more easily exposed to novel, yet unforeseen, types of attacks and misuse in the future. Our present study is framed in the context of \emph{celebrity} voices (due to \textcolor{\revcolor}{the} adoption of the VoxCeleb corpus) but we intend it as a proxy to address a specific risk associated with anyone having large quantities of biometric data in a public domain, often referred to as \emph{found data}. \textcolor{\revcolor}{In specific, we carry out empirical assessment of attacks on voice biometric system with the help of found audio data. This type of attacks have received surprisingly little attention in the literature.}
Unlike the use of publicly available tools for voice cloning of a specific target, we look for a speaker with the most similar voice and use him/her as an imposter. We use target speaker's publicly available voice \emph{data} and publicly available ASV tool for the voice similarity search.

The potential threat \textcolor{\revcolor}{of} natural impersonated voice, also known as mimicry~\cite{Gonzalez2015-mimicry}, \textcolor{\revcolor}{has been studied in a limited number of target speakers and mimickers \cite{Luck1969-cepstral,Lau-Vulnerability2004,Gonzalez2015-mimicry,Zetterholm2004-PerceptionSpeakerVerification}}. The present work is related to the study on the impact of the voice impersonation in ASV where the impersonator and potential target speakers are selected from large set of speakers. This enables us to choose the \textcolor{\revcolor}{those} impersonator-target pairs who are already similar in their natural voice. Surprisingly, the studies involving the search of potential attackers and the assessment of their ability to break the biometric security system are very limited. \textcolor{\revcolor}{For other behavioral biometric traits (than voice), perhaps the only related study is done with \emph{shoulder surfing} attack in the context of touch input implicit authentication~\cite{khan2016targeted}. This demonstrated that when potential attackers are selected and trained to perform targeted mimicry, this authentication method is highly prone to such attacks.} 

The closest prior work in spirit to our study is~\cite{Lau2005-mimicry} where the authors studied the effect of mimicry in ASV with two professional imitators and four non-professional imitators.
The closest speaker for each imitator was chosen from YOHO corpus of 138 speakers using Gaussian mixture model (GMM) based likelihood. \textcolor{\revcolor}{The study indicated that, when mimicking the most similar speaker, the professionals did not achieve better mimicry performance than non-professional imitators.} On the other hand, the professional imitators were more successful at mimicry when the target speaker is different from the most similar speaker. In another study crowdsourcing is used to select the best imitator for a set of 53 target speakers~\cite{Panjwani2014-crowd}. The authors used GMM-based ASV system for finding the imitators from a set of 176 participants. As a first step, the participants were asked to speak in natural and mimicked \textcolor{\revcolor}{voices}. Then an ASV system \textcolor{\revcolor}{was} used to filter the candidates by assessing the closeness of their voice samples to the target speakers. Finally, a set of good imitators \textcolor{\revcolor}{were} confirmed based on the performance of filtered candidates on multiple imitation tasks.  

In contrast to the studies in~\cite{Lau-Vulnerability2004,Lau2005-mimicry,Panjwani2014-crowd} with limited number of target speakers (and use of a single ASV system only), the current work uses two large publicly available datasets, VoxCeleb1 and VoxCeleb2, consisting of more than 7,000 speakers to search the targets corresponding to the six recruited participants who are native Finnish speakers. In addition to the \textcolor{\revcolor}{impersonator-specific} closest, median, and furthest targets, we also consider a common celebrity target. This is \textcolor{\revcolor}{to} evaluate the impersonator's natural ability to mimic a known person. Further, the target speakers are chosen from both Finnish and non-Finnish speakers to assess impersonator's success rate for native and non-native targets.

\section{ASV-assisted mimicry attacks}

\subsection{Attack implementation}

Let $\mathcal{T}=\{T_j\}_{j=1}^J$ denote a set of unique, publicly known \textbf{target speaker} identities and let $\mathcal{A}=\{A_k\}_{k=1}^K$ denote a set of \textbf{attacker} identities. The aim of an attacker $A \in \mathcal{A}$ is to masquerade him/herself as a specific target $T \in \mathcal{T}$ that he\textcolor{\revcolor}{/she} pre-selects using automatic speaker recognition technology. We assume that $J \gg K$ --- that is, an attacker is relatively infrequent, but there are many natural persons who have their voice samples available in a public domain. Celebrities and anyone actively uploading or streaming their video or voice data to social media platforms are representative examples.  

Given a pair of speech utterances (or a pair of \emph{collections} of multiple utterances), $(U_i,U_j)$, an \textbf{automatic speaker verification} (ASV) system (speaker detector), $\mathcal{D}(U_i,U_j)$ computes \textcolor{\revcolor}{a} \emph{detection score}, $s_{ij} \in \mathbb{R}$, typically a \emph{log-likelihood ratio} (LLR), 
    \begin{equation}
        s_{ij} = \log \frac{p(U_i,U_j|H_0)}{p(U_i,U_j|H_1)},
    \end{equation}
where the null hypothesis $H_0$ states that $U_i$ and $U_j$ originate from the same speaker and its complement $H_1$ states they originate from two different speakers. In this work, utterances are represented as fixed-sized \emph{embeddings} using either \emph{identity vectors} (i-vectors) \citep{Dehak2011-front-end} or \emph{x-vectors} \citep{snyder2018xvector}. If either $U_i$ or $U_j$ consist of multiple utterances, their embeddings are averaged. The LLR computation uses \emph{probabilistic linear discriminant analysis} (PLDA) \cite{PrinceElder2007-PLDA} scoring. The higher the LLR score, the stronger the support for the null hypothesis. %
We consider two different types of ASV systems. The first one, \textbf{attacker's ASV}  ($\mathcal{D}_\text{pub}$),
is a public-domain ASV implementation while the latter, \textbf{black-box ASV} ($\mathcal{D}_\text{black} \neq \mathcal{D}_\text{pub}$), is the system which the attacker attempts to hack into as a specific target. The attacker does not have access to the internal workings of $\mathcal{D}_\text{black}$ or its outputs to optimize mimicry attacks. 
The attack proceeds as follows:

\begin{mdframed}[style=MyFrame]\label{asv-assisted-attack}
\center{\textbf{ASV-assisted target speaker selection for mimicry attack}}
\begin{small}
    \begin{enumerate}
    \vspace{-0.1cm}
        \item Attacker $A \in \mathcal{A}$ records his/her natural voice sample, $\mathcal{U}_\text{nat}$ (one or several utterances).
        \item $A$ uses $\mathcal{D}_\text{pub}$ to
        compute scores $\{s_j\}_{j=1}^J$ between $\mathcal{U}_\text{nat}$ and all the targets in a public domain. $A$ picks the \textbf{closest target}, $j^*=\arg\max_{j=1}^J \mathcal{D}_\text{pub}(U_\text{nat},U_j)$, where $U_j$ contains all the public recordings of target $T_j$.
        \item $A$ further uses $\mathcal{D}_\text{pub}$ to pick the top-scoring utterances of $T_{j^*}$ similarly.
        \item $A$ listens to the selected utterance(s) and tries to adjust his/her voice towards the target. Once completed practicing, $A$ submits a mimicked test utterance $U_\text{mimic}$ to $\mathcal{D}_\text{black}(U_\text{mimic},U_{j^*})$ with identity claim $T_{j^*}$ (aiming to be accepted as $T_{j^*}$).
    \vspace{-0.1cm}
    \end{enumerate}
\end{small}
\end{mdframed}

\textcolor{\revcolor}{Note that in our model, the attacker uses the public-domain ASV system only to select the target speakers. In some prior work, such as \citep{Zetterholm2004-PerceptionSpeakerVerification}, ASV score was provided as feedback for the impersonators to improve their mimicry skills. We do not provide ASV (or other) feedback signals to our attackers. The main reason is that the ASV score is not necessarily intuitive to humans. For instance, a low attacker-to-target ASV score does not suggest \emph{how} to modify one's voice production so as to improve the score. Providing intuitive feedback, for instance in terms of suggested articulatory or voice source modifications, would require a different system (and user interface) design. In our model, the attacker uses a readily-available public-domain ASV system to rank and select potential target speakers, but \emph{without} any further numerical feedback or system optimization. Such `passive' ASV system could be, for instance, a voice search service that finds most similar speakers to the user's voice from a public video archive --- see \cite{Vestman2019-who-do,Celebsoundalike} as examples.} 

\textcolor{\revcolor}{Both the attacker's and the attacked ASV systems are \emph{text-independent}, \emph{i.e.} none assumes the spoken contents of the compared enrollment and test utterances to match. Even if properly-optimized text-dependent ASV systems can provide higher recognition accuracy, text-independent ASV systems provide more flexibility and are justifiable in certain authentication applications, such as secure teleconferencing and telephone banking. The use of text-independent ASV systems in this study was, in fact, \emph{necessary} as we have no control over the text content in the celebrity corpus (VoxCeleb).} 

\subsection{Public-domain (attacker's) ASV system}

The attacker's ASV \textcolor{\revcolor}{system} uses i-vector front-end \citep{Dehak2011-front-end} and probabilistic discriminant analysis (PLDA) \citep{PrinceElder2007-PLDA} back-end to compute speaker similarity scores. \textcolor{\revcolor}{The system's acoustic front-end\footnote{\url{http://cs.joensuu.fi/~sahid/codes/AntiSpoofing_Features.zip}} extracts} 20 mel-frequency cepstral coefficients (MFCCs) \textcolor{\revcolor}{per frame} using 20 filters, leading to 60 features per frame after including deltas and double-deltas. \textcolor{\revcolor}{The chosen MFCC configuration is commonly used in speaker recognition experiments~\cite{ALAM2013237,Dehak2011-front-end}.} The features are processed with RASTA filtering \citep{Hermansky94-RASTA} and cepstral mean and variance normalization (CMVN). Non-speech frames are omitted using energy-based speech activity detector (\textcolor{\revcolor}{SAD}) (described in Section 5.1 of~\citep{kinnunen2010overview}).

\textcolor{\revcolor}{
The universal background model (UBM), i-vector extractor, linear discriminant analyzer (LDA), and PLDA, are trained using Wall Street Journal (WSJ) and Librispeech corpora. LDA is used to reduce 400-dimensional i-vectors to 250 dimensions before centering, whitening, and length normalization. Simplified PLDA with 200-dimensional speaker subspace is used for scoring. For further details, refer to Table \ref{table:asv-systems} of the current work and Section~2.2 of \cite{Kinnunen2018-can-we}.}

\renewcommand{\arraystretch}{0.75}
\begin{table*}[t]
\caption{Details of the speaker verification systems used to simulate targeted impersonation attack against automatic speaker verification. The attacker is assumed to not have information about the attacked \textcolor{\revcolor}{system}, and hence the attacker's system differs from the attacked \textcolor{\revcolor}{system}.}
\label{table:asv-systems}
\footnotesize
\centering
\makebox[1.0\linewidth]{\begin{tabular}{p{0.23\linewidth-2\tabcolsep} p{0.42\linewidth-2\tabcolsep} p{0.42\linewidth-2\tabcolsep}}
\hline
& \textbf{Attacker's ASV system} & \textbf{Attacked ASV system}\\
& $(\mathcal{D}_\text{pub})$ & $(\mathcal{D}_{\text{black}})$ \\
\hline\hline

\textcolor{\revcolor}{Type} & \textcolor{\revcolor}{Text-independent} & \textcolor{\revcolor}{Text-independent}\\[2ex]

Implementation & MSR Identity Toolkit (MATLAB) & Kaldi (c++) \\[2ex]

Sampling rate & 16 kHz & 16 kHz\\[2ex]

Acoustic features & 60 MFCCs (20 static+20-$\Delta$+20-$\Delta\Delta$), RASTA, SAD, CMVN & 30 MFCCs (no deltas), Sliding CMN normalization, \textcolor{\revcolor}{SAD}\\[2ex]

Embedding type & i-vector (400-D) & x-vector (512-D)\\[2ex]

Back-end / scoring & LDA (250-D)+PLDA (simplified, 200-D) & LDA (200-D)+PLDA (2-cov)\\[2ex]

Development data & Librispeech (train-clean-360 and train-clean-100 subsets), WSJ0 and WSJ1  & VoxCeleb2, training part of VoxCeleb1\\[2ex]

Data augmentation & None & Reverberation, noise, music, babble\\[2ex]

EER* & 12.84 (\%) & 3.11 (\%) \\
\hline
\multicolumn{3}{r}{} \textcolor{\revcolor}{* \footnotesize{EER for VoxCeleb1 test protocol}}
\end{tabular}}
\end{table*}
\renewcommand{\arraystretch}{1.0}

\subsection{Attacked ASV \textcolor{\revcolor}{system}}

In our experiments, we regard \textcolor{\revcolor}{the} x-vector system \citep{snyder2018xvector}, based on pre-trained Kaldi \citep{Povey_ASRU2011} \textcolor{\revcolor}{recipe}, as \textcolor{\revcolor}{the} ASV system to be attacked. To emulate the scenario of attacker's limited knowledge of this system,
the attacker's ASV is made intentionally different from the attacked ASV \textcolor{\revcolor}{system} in terms of feature extractor set-up, embedding type, and development corpora (Table \ref{table:asv-systems}). \textcolor{\revcolor}{The} attacked \textcolor{\revcolor}{system} is \textcolor{\revcolor}{the} Kaldi x-vector \textcolor{\revcolor}{recipe} for VoxCeleb, while the attacker's system uses i-vectors. Unlike \textcolor{\revcolor}{the} i-vector \textcolor{\revcolor}{extractor}, \textcolor{\revcolor}{the} x-vector \textcolor{\revcolor}{extractor is} trained discriminatively using speaker labels.

\vspace{-0.2cm}
\section{Corpus of target speakers: VoxCeleb}
\vspace{-0.2cm}
The attacker's ASV is used as a voice search tool to find the closest speakers from the combination of VoxCeleb1 \citep{Nagrani2017} and Voxceleb2 \citep{Chung2018} to each of the locally recruited subjects (described in Section \ref{sec:local-attackers}). The combined VoxCeleb corpus contains about $1.3$ million speech excerpts extracted from more than 170,000 YouTube videos from $J=7,365$ unique speakers. This totals to about 2,800 hours of audio material, most of which is active speech. Both VoxCeleb corpora were collected using automated pipeline exploiting face verification and active speaker verification technologies \citep{Chung2018}.

VoxCeleb1 contains mostly English speech, while VoxCeleb2 is more diverse in nationalities and languages. The nationality information of the target speakers was of our interest, as the recruited local speakers are Finnish and we wanted to see if Finnish people do better job at imitating \textcolor{\revcolor}{Finnish rather than} non-Finnish targets. According to the VoxCeleb1 metadata, there are no Finnish speakers in VoxCeleb1. \textcolor{\revcolor}{VoxCeleb2 did not include} nationality metadata but we extracted the nationalities automatically using Google's \emph{Knowledge Graph} API\footnote{\url{https://developers.google.com/knowledge-graph/}}. This way we identified a total of 44 Finnish speakers from VoxCeleb2.
\vspace{-0.3cm}
\section{Locally recruited attackers}\label{sec:local-attackers}

\vspace{-0.1cm}
\subsection{Speakers and recording gear}
\vspace{-0.1cm}
We recruited $K=6$ voluntary local speakers (4M + 2F) to serve as `attackers'. The selected terminology, `attacker', is made for convenience to reflect the focus of ASV vulnerability study; it should be understood that all speakers took part voluntarily and were not asked to `hack' any computer systems in the sense understood in the security field. In fact, most of our speakers are considered \emph{naive} to the study aims: two of the male subjects knew the specific goals of the study but the remaining four subjects were not informed that the text and target speakers were tailored for them, nor where the target voices were obtained from. The speakers were not informed that the study relates to ASV vulnerability, but were asked to mimic the target speakers as accurately as they could. All the subjects signed an informed consent form to use their speech data for research, and were rewarded with movie and coffee tickets. 

\textcolor{\revcolor}{
All six attackers are native Finnish speakers with an age range between 24 to 44 years old. They are \emph{naive} impersonators who lack formal training in mimicry.
We adopt the same recording setup from  \citep{GonzalezHautamaki2017-acoustical} and text prompts are described in detail in \cite{CSL2018_additional_material}}.
As illustrated in Fig. \ref{fig:study-overview}, the subjects took part to three recording sessions. The first session, produced in the subject's natural voice, is used for VoxCeleb target speaker selection, while the remaining two sessions serve for vulnerability analysis of the attacked systems. The tasks in the recording sessions differed, while the recording set-up was the same: recordings took place in a silent laboratory room with a portable Zoom H6 Handy Recorder using an omnidirectional headset mic (Glottal Enterprises M80) with 44.1 kHz sampling and 16-bit quantization. Three other channels (two smartphones and electroglottograph) were also collected, but are not used in this study.

\subsection{The first recording session (data for target search)}

\label{sec:first_recording_session}

The first session, used for the targeted VoxCeleb speaker search, consists of four tasks in the speaker's natural voice. The tasks consisted of spontaneous speech and read text (13 sentences) in both Finnish and English. The read texts in Finnish are the same used in \citep{GonzalezHautamaki2017-acoustical}. Their corresponding English versions were added for this study. We have approximately six minutes of speech (before speech activity detection) per speaker from Session 1. \textcolor{\revcolor}{Detailed description of the material used in data collection can be found in the online supplementary material \cite{CSL2018_additional_material}.}

\vspace{-0.1cm}
\subsection{Attacked target speaker search and utterance selection}
\vspace{-0.1cm}
\label{sec:selection}
For the purpose of targeted speaker search, we compute a single averaged i-vector for each of the six speakers resulting from 28 individual utterances from Session 1.
Similar to \citep{Lau-Vulnerability2004}, we use the ASV system to pick for each attacker the \textbf{closest}, \textbf{median}, and \textbf{furthest} speakers among the VoxCeleb speakers. The closest one is most relevant for vulnerability analysis while the other two serve for reference purposes. We do this ASV-assisted search separately for \emph{all} the VoxCeleb speakers (unconstrained search from 7,365 speakers) and for the subset of 44 Finnish speakers. We pool all the speech data of the VoxCeleb speakers to compute average i-vector per target. \textcolor{\revcolor}{The selected target speakers per attacker are presented in Tables~\ref{tab:fi_targets} and \ref{tab:nonfi_targets}.}
\renewcommand{\arraystretch}{0.7}
\begin{table}[htbp]
\caption{Target speakers (closest, median and furthest) per attacker. Selection of potential targets from 44 Finnish celebrities in VoxCeleb2.}
\label{tab:fi_targets}
\small
\begin{tabular}{|c|l|l|l|}
\hline
\textbf{Attacker ID } & \textbf{Celebrity} & \textbf{Profession} & \textbf{Spoken language} \\ \hline
M1 & Samuli Edelmann & Actor, singer & Finnish, English \\ 
 & Paavo V\"ayrynen & Politician & Finnish \\ 
 & Antti Tuisku & Pop singer & Finnish \\ \hline
M2 & Samuli Edelmann & Actor, singer & Finnish, English \\ 
 & Paavo V\"ayrynen & Politician & Finnish \\ 
 & Mika Kojonkoski & Ski jumper, politician & Finnish, English \\ \hline
M3 & Joni Ortio & Ice hockey player & Finnish, English \\ 
 & Elastinen & Rap musician & Finnish \\ 
 & Perttu Kivilaakso & Musician & English \\ \hline
M4 & Samuli Edelmann & Actor, singer & Finnish, English \\ 
 & Tuomas Holopainen & Musician & Finnish, English \\ 
 & Jyrki Katainen & Politician & Finnish, English \\ \hline
F1 & Anna Puu & Pop singer & Finnish \\ 
 & Karita Mattila & Opera singer & Finnish, English \\ 
 & Tarja Halonen & Politician & Finnish, English \\ \hline
F2 & Sofi Oksanen & Writer & Finnish, English \\ 
 & Kaisa M\"ak\"a\"ar\"ainen & Biathlete & Finnish, English \\ 
 & Tarja Halonen & Politician & Finnish, English \\ \hline
\end{tabular}
\end{table}

\begin{table}[htbp]
\caption{English speaking celebrities (closest, median and furthest) per attacker. Selection from 7321 potential targets in VoxCeleb1 and VoxCeleb2. * indicates speakers from VoxCeleb1.}
\label{tab:nonfi_targets}
\small
\begin{tabular}{|c|l|l|l|}
\hline
\textbf{Attacker ID } & \textbf{Celebrity} & \textbf{Profession} & \textbf{Spoken language} \\ \hline
M1 & Valentin Inzko & Politician & English (Austrian) \\ 
 & Elijah Cummings & Politician & American English \\ 
 & Chris Colfer  * & Actor & American English \\ \hline
M2 & Jeremy Irons * & Actor & British English \\ 
 & Karan Tacker & Actor & Indian English \\ 
 & Ryan Ochoa * & Actor & American English \\ \hline
M3 & \'Eric Boullier & F1 manager & English (French) \\ 
 & Guillaume Canet * & Actor, director & English (French) \\
 & Bill Gilman & Singer & American English \\ \hline
M4 & Ciar\'an Hinds & Actor & Irish English \\ 
 & Ian Kinsler  & Baseball player & American English \\ 
 & Phil Mickelson & Golf player & American English \\ \hline
F1 & Jessie J * & Singer & British English \\ 
 & Candace Cameron * & Actress & American English \\ 
 & Lin Shaye * & Actress & American English \\ \hline
F2 & Fay Ripley & Actress, author & American English \\ 
 & Belcim Bilgin & Actress & English (Turkish) \\ 
 & Anne Hathaway * & Actress & American English \\ \hline
\end{tabular}
\end{table}
\renewcommand{\arraystretch}{1.0}

In addition to the three ASV-selected targets, we include \textbf{common target} matched with the speaker's gender, in both Finnish and English. The common Finnish speaking targets are P{\"a}ivi R{\"a}s{\"a}nen (female, politician) and Ilkka Kanerva (male, politician), and the common English speaking targets are Hillary R. Clinton (female, politician) and Leonardo DiCaprio (male, actor). \textcolor{\revcolor}{The choice of the common targets is arbitrary but based on a loose, subjective criterion \emph{as famous as possible}.
We first identified a short-list of VoxCeleb celebrities that we thought are well-known. We then ran an e-mail survey among our friends and colleagues (23 responded), asking each one to indicate the three most famous persons (in their opinion). We combined their votes to select the common targets.}
Even if the selected targets are well-known, from the viewpoint of ASV they are \emph{random} target speakers with no strong presuppositions how similar their voices are to our attackers. 

In summary, for each of our \textcolor{\revcolor}{four} male and \textcolor{\revcolor}{two} female subjects, we select \textcolor{\revcolor}{six} customized targets (\textcolor{\revcolor}{three} ASV-ranks $\times$ \textcolor{\revcolor}{two} languages) and \textcolor{\revcolor}{two} common gender-matched ones (one Finnish, one English). This gives a theoretical total of $3 \times 2 \times 4 \,\text{male} + 2\,\, \text{common male}$ + $3 \times 2 \times 2\,\, \text{female} + 2\,\, \text{common female} = 40$ target speakers. \textcolor{\revcolor}{But as the reader can see from Table \ref{tab:fi_targets}}, not all of the ASV-selected targets are unique: one Finnish male celebrity \textcolor{\revcolor}{(Edelmann)} was the closest target for three attackers, one Finnish male celebrity repeated as the median speaker for two male attackers \textcolor{\revcolor}{(V\"ayrynen)}, and one Finnish female celebrity \textcolor{\revcolor}{(Halonen)} is the furthest speaker for both female attackers. \textcolor{\revcolor}{These collisions might be explained by the the limited number of Finnish celebrities (30M, 14F) in VoxCeleb.}
The total number of unique celebrity targets is 36.

For each of the 36 target speakers, we selected \textcolor{\revcolor}{multiple short utterances so that, when combined, each target would have} at minimum 30 seconds of active speech. \textcolor{\revcolor}{The selected utterances were used to evaluate} the ASV system attacks. \textcolor{\revcolor}{We selected only short utterances} for two reasons. First, the duration of \textcolor{\revcolor}{most of the} VoxCeleb excerpts varies between \textcolor{\revcolor}{five} to \textcolor{\revcolor}{ten} seconds. Second, we deemed shorter utterances to be easier for our attackers to imitate. \textcolor{\revcolor}{Detailed description of these utterances is provided in an online supplementary material \cite{CSL2018_additional_material}}. 

\textcolor{\revcolor}{The selection of the VoxCeleb excerpts was done by utilizing attacker's ASV system. For the closest and furthest targets we selected, respectively, the highest and lowest scoring utterances. For the median speakers, we selected the utterances closest to the mean.} This was further accompanied by manual inspection: if the audio quality (determined subjectively by listening) in a given utterance was not deemed high enough, we discarded it and moved on to the next ones in the ranked list.

\subsection{Speech transcription and the mimicry recordings}

Unlike the first recording session (common to all subjects), the second and third sessions were tailored for each subject. This process involved the use of speech transcripts of the selected target utterances. To this end, we used Amazon's Mechanical Turk\footnote{\url{https://www.mturk.com/}} (MTurk), a commercial crowdsourcing service, to transcribe the English language audio. The Finnish transcripts were produced by two native Finnish speakers. The 35 MTurk crowdworkers and the \textcolor{\revcolor}{two} Finnish transcribers were asked to transcribe all the nuances of conversational speech, including repetitions, hesitations, filler words \emph{etc}. Finally, two reviewers audited the quality of all the transcripts. \textcolor{\revcolor}{All the final transcriptions are provided in the supplementary material \cite{CSL2018_additional_material}.}

In Session 2, which took place \textcolor{\revcolor}{five} to \textcolor{\revcolor}{six} weeks after Session 1, the subject was provided with the  transcripts of the selected target utterance(s) and was asked to read the sentences twice in his or her natural voice. The speaker was not informed whose speech the transcripts corresponded to.
The rationale of including this session was to familiarize each attacker with the target speaker sentences. We adopted the general idea to include a session with reference text only and another one with audio from the design used in \citep{mariethoz2005-professional}. \textcolor{\revcolor}{In that study, the target speakers were public personalities that each impersonator knew. Each impersonator completed three scenarios with an increasing level of detail about the target speakers. The impersonator was first asked to produce prototypical target speech without knowledge of text (other than common 
category, \emph{e.g.} everyday sentences). The impersonator was then revealed the target speaker texts to be impersonated and, finally, he would be provided audio reference of target.}

In the last session, which took place \textcolor{\revcolor}{two} to \textcolor{\revcolor}{six} days after Session 2, the subjects were provided with the same transcript as in Session 2. Additionally, they were now provided access to the actual target speaker audio excerpts. The transcripts were provided on a printed paper and the audio was presented through headphones connected to a tablet computer with an interactive webpage. The subject was allowed to interact with the audio samples and could listen to the target utterance(s) as many times as needed, and he/she then tried to mimic the voice according to their best skills. Again, the subject was asked to mimic each sentence twice. In the experiments, we use only the second recording of each sentence.

\textcolor{\revcolor}{Following standard convention in the context of spoofing and countermeasure studies \cite{Wu2015-spoofing-survey}, we refer to the speech recordings of the second session as \emph{zero-effort}. This is to signify that the attackers were instructed to produce target speaker texts in their own modal voice, \emph{i.e.} without dedicated effort to sound like the target. The recordings from the last session, in turn, are simply referred to as \emph{mimicry} utterances.}

\section{Results: mimicry attacks against automatic verification system}
\label{sec:asv_results}

\begin{figure*}[t!]
\centerline{\includegraphics[width=1.0\linewidth]{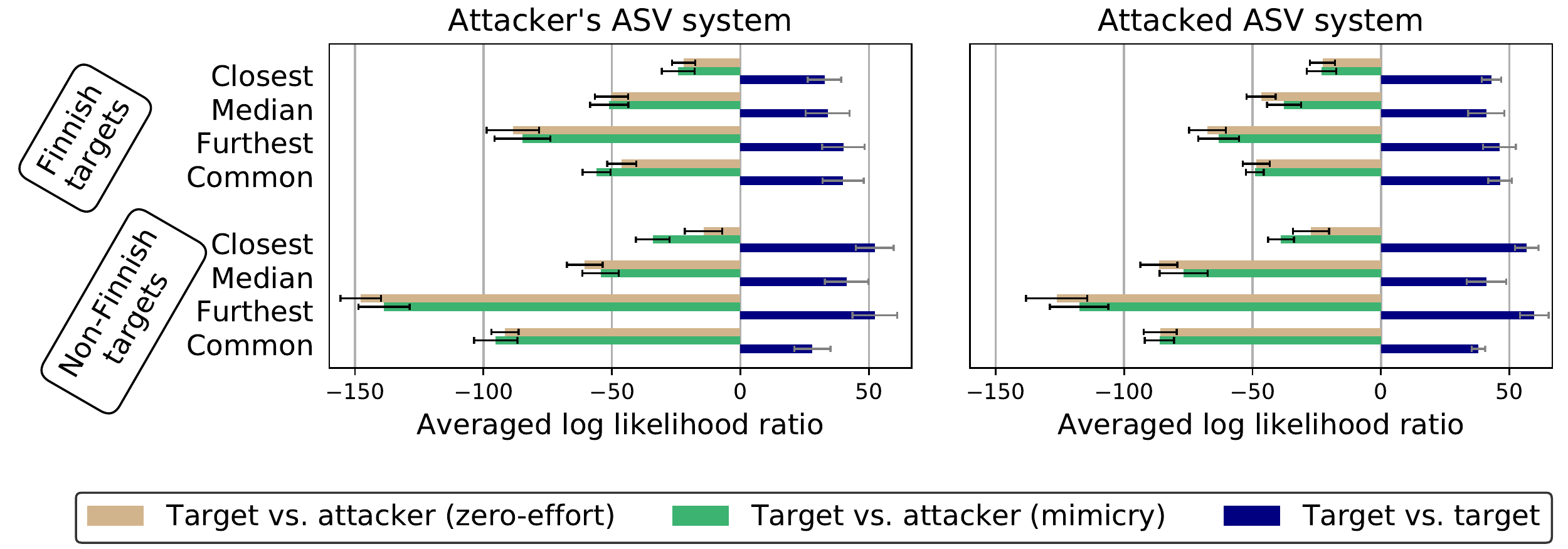}}
\caption{Comparison of attackers' ASV scores (log likelihood ratios) to the targets' scores for \textcolor{\revcolor}{both of the} ASV systems involved in the study. The scores are averaged over all attackers and all speech segments. The error bars represent 95 \% confidence intervals for the means.}
\label{fig:res}
\end{figure*}

In the following, we evaluate \textcolor{\revcolor}{the} effectiveness of \textcolor{\revcolor}{mimicry attacks} against ASV systems. The target speaker models used in the experiments were enrolled using all available segments except those selected for testing as described in Section~\ref{sec:selection}.

Figure \ref{fig:res} displays how the PLDA scores of genuine and attack trials compare to each other. The general findings are as expected. First, the order of the closest, the median, and the furthest speakers transfers from the attacker's ASV system to the attacked ASV system, implying that the ASV-assisted speaker selection \emph{can} help in ASV attacks. Second, in general, the \textcolor{\revcolor}{attackers'} natural and mimicry scores are significantly (by a wide margin) below the target scores. Additionally, we find no significant difference between the \textcolor{\revcolor}{zero-effort and mimicry attacks (except for the closest category).} Finally, as the recruited attackers \textcolor{\revcolor}{are} Finnish, attackers' scores against the Finnish targets \textcolor{\revcolor}{are} higher than for the non-Finnish targets (within each rank category).

\textcolor{\revcolor}{We further display the difference of mimicked and natural speech scores in Table \ref{table:mimic}}. Interestingly, and contradictory to what we assumed, if the target speaker's voice is already close to the attacker's voice, the impersonation attempts \emph{degrade} the score. The same finding was noted in situations where the target \textcolor{\revcolor}{is a well known public figure (as the targets in the common category are)}. We suspect that the effect might be due to people having higher tendency to \emph{overact} someone they already know well. However, if the targets are not close to the attackers (\emph{i.e.}, median and furthest categories) or are less well known, impersonation is potentially helpful (though, not by a statistically significant margin).

\renewcommand{\arraystretch}{1.0}
\begin{table}[t]
\caption{Score differences between attacks with impersonated voices and attacks with natural voices. Differences are averaged over attackers, target nationalities, and utterances. $\pm$ indicates 95 \% confidence intervals. In the case of the closest target speakers, impersonation attempts are counterproductive.}
\vspace{2mm}
\label{table:mimic}
\small
\begin{tabular}{p{0.20\linewidth-2\tabcolsep}>{\raggedleft\arraybackslash}p{0.20\linewidth-2\tabcolsep}>{\raggedleft\arraybackslash}p{0.21\linewidth-2\tabcolsep}>{\raggedleft\arraybackslash}p{0.19\linewidth-2\tabcolsep}>{\raggedleft\arraybackslash}p{0.20\linewidth-2\tabcolsep}}
\hline
\mbox{ASV system} & Closest & Median & Furthest & Common \\
\hline
Attacker's ASV & -9.7 $\pm$ 5.2 & 2.2 $\pm$ 4.3 & 5.9 $\pm$ 7.1 & -7.2 $\pm$ 4.3\\
Attacked ASV & -5.2 $\pm$ 3.9 & 9.2 $\pm$ 3.3 & 6.1 $\pm$ 4.3 & -0.5 $\pm$ 3.8\\
\hline
\end{tabular}
\end{table}
\renewcommand{\arraystretch}{1.0}

\begin{figure}[h]
\centerline{\includegraphics[width=1.3\textwidth]{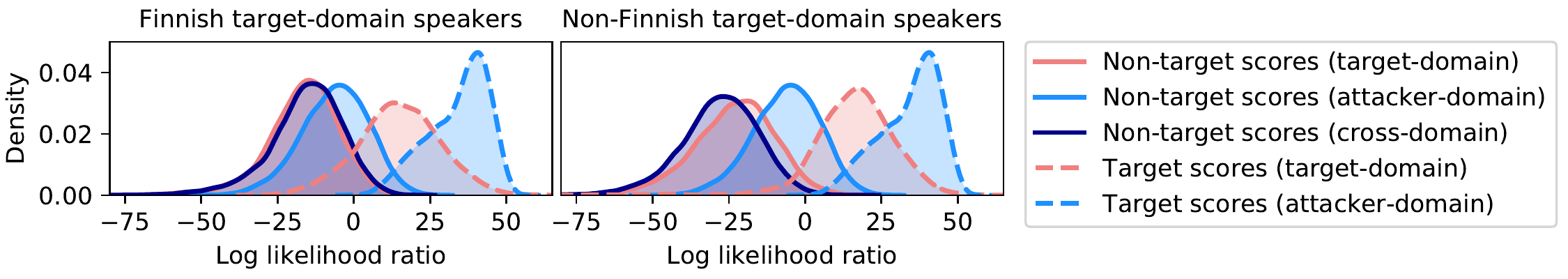}}
\caption{\textcolor{\revcolor}{Distributions of target and non-target scores in different domains. \emph{Cross-domain} non-target scores are obtained by scoring speakers from the attacker domain against the speakers from the target (VoxCeleb) domain. The simulated mimicry attacks in this work fall under the category of cross-domain trials. As the cross-domain score distributions overlap almost perfectly with the target-domain non-target distributions, the domain mismatch does not seem to make attacking more difficult, at least when the targets are Finnish.}}
\label{fig:score_distributions}
\end{figure}

Our attackers are native Finnish speakers recorded with a specific set-up which may differ from the \textcolor{\revcolor}{target domain (VoxCeleb)} conditions. 
\textcolor{\revcolor}{
This raises a question whether our mimicry attacks might have been unsuccessful due to \emph{domain mismatch}.
To address this question, we studied \emph{target-domain}, \emph{attacker-domain}, and \emph{cross-domain} non-target score distributions as well as target-domain and attacker-domain target score distributions. It was not possible to construct cross-domain target trials as we do not have speakers common to both domains. The main interest in this specific study is to compare target-domain non-target scores to cross-domain non-target scores. If the cross-domain scores (the case of attacks) do not fall below the target-domain scores, it suggests that the attacker does not get penalized by the domain mismatch. The scores for the study were obtained from the attacked x-vector based ASV system.}

\textcolor{\revcolor}{
Figure \ref{fig:score_distributions} indicates that when the nationality mismatch is present (non-Finnish target-domain speakers), the cross-domain non-target scores are, on average, slightly lower than the the target-domain non-target scores. If, however, the target-domain speakers are Finnish, like our recruited attackers are, the non-target speaker distributions overlap almost perfectly. This suggests that the Finnish attackers attacking the Finnish VoxCeleb targets did not seem to get penalized by the domain mismatch. The domain mismatch can be observed by comparing target and non-target scores of attacker-domain and target-domain. As the attacker-domain is has much less variability in the conditions, the scores in attacker-domain tend to be higher.
}

\section{Perceptual evaluation of mimicry attacks}\label{sec:perceptual-evaluation}

Next, we evaluated how ASV assisted mimicry attacks perform against human listeners. Further, we compared the findings of perceptual test to those obtained from the attacks against the ASV system. \textcolor{\revcolor}{To avoid nationality mismatch between targets and attackers, we restricted our experiments to Finnish targets only.}

\subsection{Listening test setup}

In total, we had \textcolor{\revcolor}{625} pairs of speech samples (trials) to be evaluated by the listeners. These trials can be divided \textcolor{\revcolor}{into} five groups of \textcolor{\revcolor}{125} trials (\textcolor{\revcolor}{4 to 7} trials for each of the 24 attacker-target combinations). The first three groups are related to the mimicry attacks: 1) target vs.\ target (reference point), 2) target vs.\ attacker (zero-effort mimicry), and 3) target vs.\ attacker (mimicry). For each set of three trials, the same target enrollment utterance is used. The speech content of the test utterances is the same in all three cases, but different from that of the enrollment utterance (\emph{i.e.} text-independent speaker comparison). The two last types of trials focus on the attacker. They are 4) attacker (zero-effort) vs.\ attacker (zero-effort) and 5) attacker (zero-effort) vs.\ attacker (mimicry). These two cases \textcolor{\revcolor}{are included, respectively, to study the listeners' performance for the same-speaker trials with fixed recording conditions, and to study how much the attackers modify their voices relative to their natural voices when mimicking.} In the cases 4) and 5), the enrollment utterances are selected from the English part of the data described in Section~\ref{sec:first_recording_session}. Similarly as above, for each set of two trials, the enrollment utterance is fixed and the two test utterances have the same content. In all of the cases, the enrollment utterance was selected from the available utterances so that its duration is close to the duration of \textcolor{\revcolor}{the} test utterances.

The listening trials were accompanied with a question \emph{``How similar the two speakers in the two voice samples sound to you?''}, to which the listeners answered using a 4-point scale with options \emph{Very dissimilar}, \emph{Dissimilar}, \emph{Similar}, and \emph{Very similar}. The 4-point scale was selected to enforce the listeners to make up their mind regarding speaker similarity. When presenting the trials, the order of the two voice samples in a trial was randomized so that the enrollment utterance was not always played the first. Each trial was presented individually and their order was randomized as well. For each of the \textcolor{\revcolor}{625} trials, we asked opinions from five different listeners, so in total we collected \textcolor{\revcolor}{3125} responses from the listeners.

\textcolor{\revcolor}{We recruited the listeners using the Amazon's MTurk service.}
All the listeners were either native English speakers or had advanced English skills. In total, \textcolor{\revcolor}{225} crowdworkers participated the listening trials. \textcolor{\revcolor}{Five} workers rated more than 100 trials, whereas \textcolor{\revcolor}{130 completed less than five. On average, a  crowdworker completed $3125/225 \approx 14$ trials. Out of the 225 listeners, 40 provided information about their mother tongue: 26 English, 4 Italian, 4 Portuguese, 2 German, 2 Spanish, 1 Estonian, 1 Tamil.}

\subsection{Listening test results}

\begin{figure}[t]
\centerline{\includegraphics[width=1.3\linewidth]{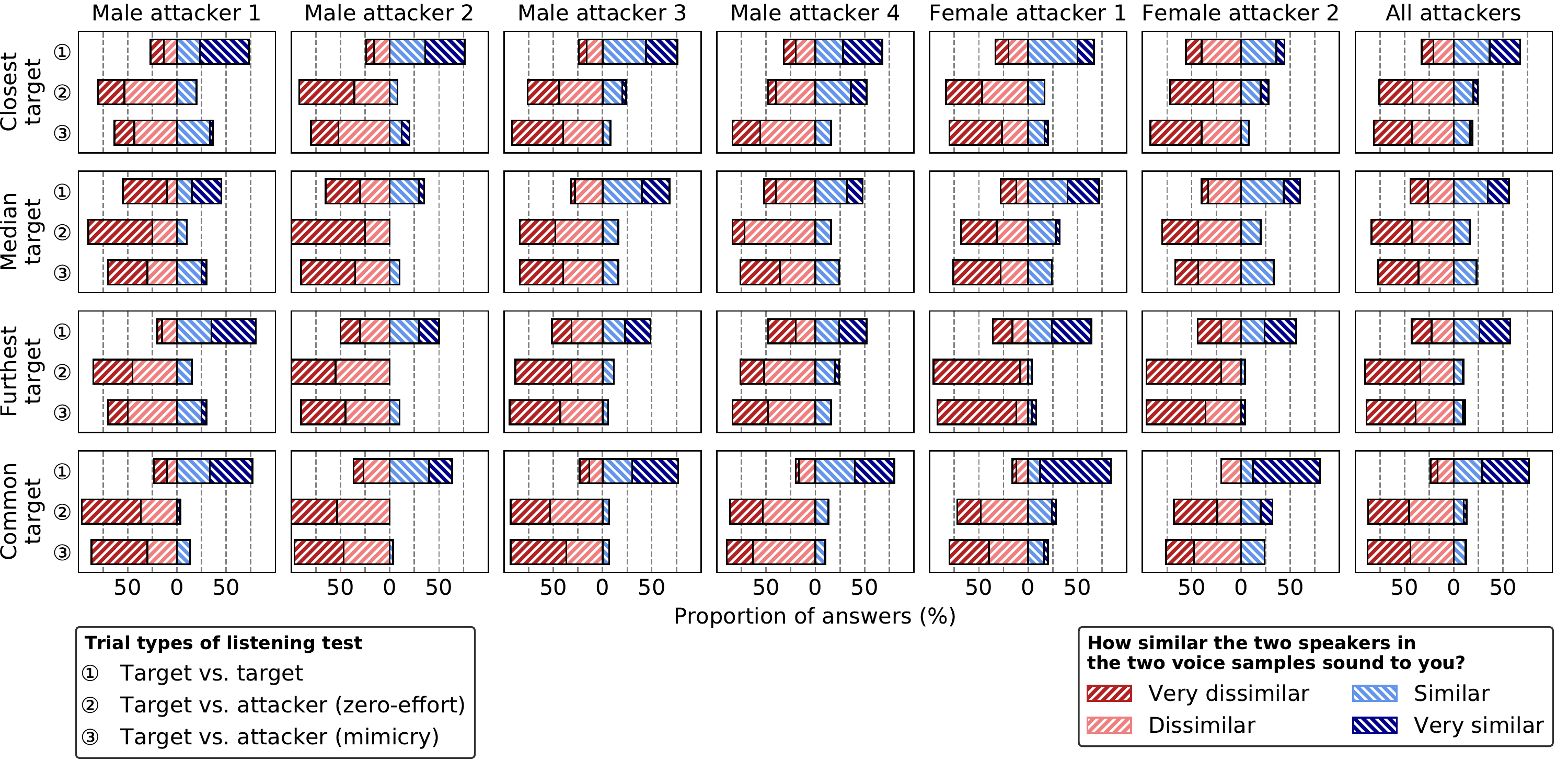}}
\caption{Results from the listening test (target speaker enrollment vs.\ test segment). Each attacker (in columns) has 4 targets speakers (in rows: closest, median, furthest, common). For each attacker-target combination, there are three different trial types (denoted by circled digits) as described in the left-hand side legend. The last column shows the results when trials from all the attackers are combined.} 
\label{fig:listener_scores}
\end{figure}

We present the main results of the listening test in Figure~\ref{fig:listener_scores}, which presents the listener judgements of speaker similarity for all the studied attacker-target combinations.
First, the listeners regard the two samples from the same target speaker (target vs.\ target cases) similar or very similar to each other, as expected. However, there are individual cases that turned out to be difficult for the listeners. For example, the \textcolor{\revcolor}{median target of the male attacker 1 was considered dissimilar or very dissimilar sounding to himself in most of the answers.} Informal listening of the utterances of this target revealed that the target's voice sounded different each time mostly due to differences in speaking style, recording conditions, and audio processing. For example, in one sample, the target speaker \textcolor{\revcolor}{(Finnish politician) is being interviewed in a talk show, whereas in another sample he is giving a public speech in very different conditions.}

How are the listeners opinions affected by mimicry? On average (see the last column of Figure~\ref{fig:listener_scores}), mimicry does not seem to help to make the attackers sound more like the targets. 
\textcolor{\revcolor}{At the individual level, we find, however, that male attackers 1 and 2 got higher ratings for their mimicked speech}.
Further, we find that ASV assisted target speaker selection can help in choosing attacker-target pairs that sound similar to each other. That is, the furthest targets get lower similarity ratings than the closest targets. Even if automatic systems and humans based their speaker similarity judgments differently, the broad rank categories seem consistent.

\begin{figure}[t]
\centerline{\includegraphics[width=1.3\linewidth]{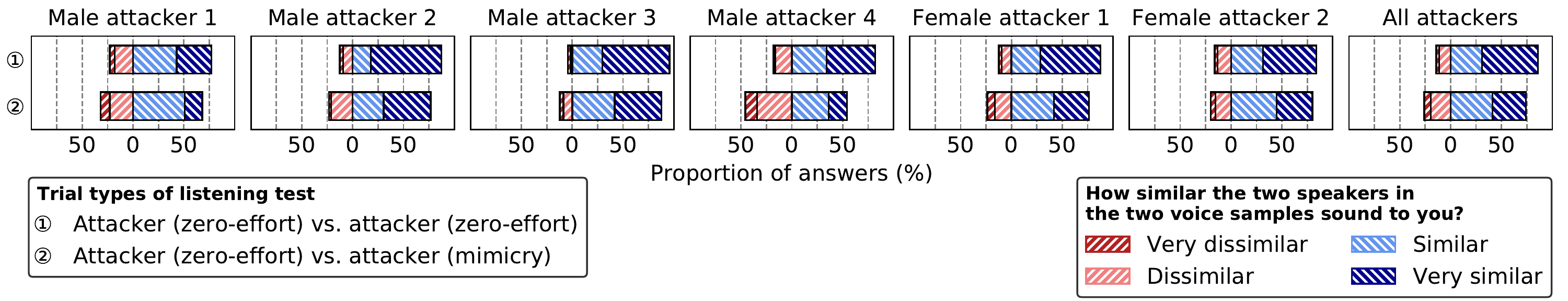}}
\caption{Results from the listening test (attacker enrollment vs.\ attacker test segment). Listeners evaluate each attacker's enrollment samples against attacker's zero-effort and mimicry-effort attack samples. The voice modification induced by mimicry attempt makes the attackers sound less like themselves.}
\label{fig:listener_scores_2}
\end{figure}

Figure~\ref{fig:listener_scores_2} displays listening test results for those trial types where attacker's enrollment utterances are compared to attacker's test segments with and without mimicry effort. The same-speaker trials have higher similarity ratings in comparison to those in Figure~\ref{fig:listener_scores}). This is expected since our attacker corpus is practically free from channel variation and background noise unlike the VoxCeleb collections. In addition, we find that when the attackers are trying to mimic the voices of the target speakers, they sound \textcolor{\revcolor}{a little bit} less like themselves.

\subsection{Comparison of human listeners and automatic speaker verification system}

\begin{figure}[t!]
\centerline{\includegraphics[width=1.3\linewidth]{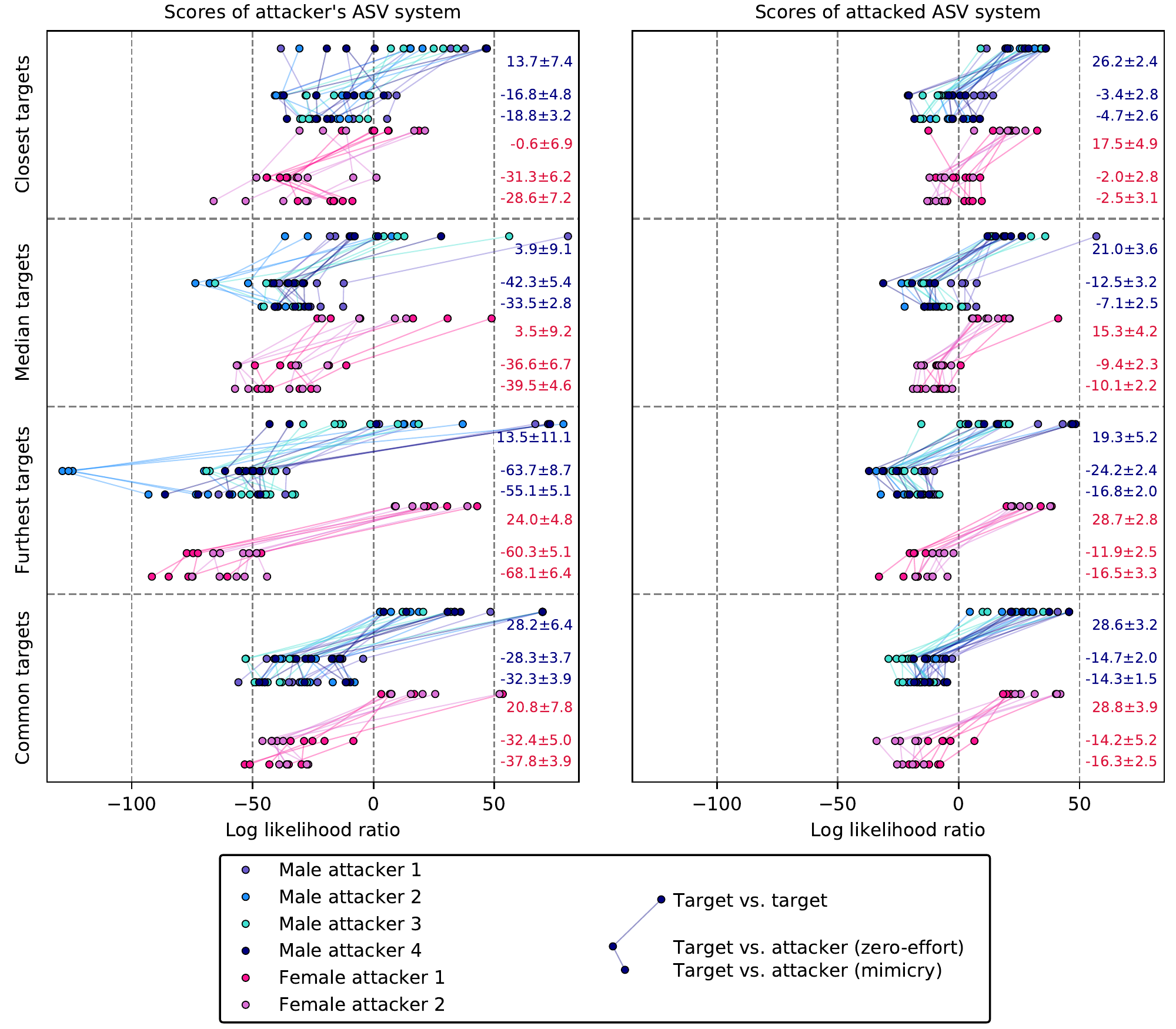}}
\caption{The scores of the ASV systems for \textcolor{\revcolor}{the} trials used in the listening test. The scores in each score triplet (described in the legend) are from the trials that have the same target speaker enrollment utterance and the speech content is the same in all the three test segments. Scores for male and female attackers are shown in separate groups. The right side of each graph displays the mean values of the score groups together with standard error of the mean multiplied by $1.96$.}
\label{fig:asv_scores}
\end{figure}

To compare human opinions to ASV system scores, we scored the same trials using both the attacker's ASV system and the attacked ASV system. All the individual scores for three different trial types are displayed in Figure \ref{fig:asv_scores}. The scores for the content matching test utterances are connected with lines and thus form score-triplets. This allows us to see how close the attacker's scores are to the target's scores and how successful were the mimicry attempts in individual cases. The results agree with the results of Figure \ref{fig:res}, as expected --- the only difference with the earlier ASV protocol is the number of target speaker enrollment utterances, which is now only one\footnote{\textcolor{\revcolor}{In general, data processing capacity of ASV systems and listeners differ: ASV systems can process multiple enrollment utterances and large number of trials, but humans have limited attention span and memory and cannot process many trials (or excessively long utterances). For the maximum benefit of the ASV system, the earlier ASV protocol used in Fig. \ref{fig:res} used multiple enrollment utterances, while the scaled-down ASV protocol (single enrollment utterance) used in Fig. \ref{fig:asv_scores} was designed to facilitate perceptual speaker comparisons.}}.

In general, the findings from the listening test are similar to what the ASV system scores imply. The ASV-assisted target speaker selection helps to bring attacker's scores closer to the target's scores, while the mimicry attempts do not seem to help much to bring the scores closer to the target's scores.

\section{Prosody \textcolor{\revcolor}{and formant} analysis of mimicry attacks}\label{sec:prosody-analysis}

To gain further insight \textcolor{\revcolor}{how attackers' change their voices to mimic their targets}, we carried out a study of the changes in fundamental frequency (F0), speaking rate, \textcolor{\revcolor}{and formants}. \textcolor{\revcolor}{Our main motivation to study these qualities is to see whether attackers changed more their prosody than spectral cues. If this is the case, the changes might not be reflected by ASV scores as our systems are based on spectral features.}

\subsection{Estimation of fundamental frequency and speech rate}

Speaking rate, in terms of syllable rate (the number of syllables per second), was measured using a Praat~\citep{Praat} implementation \citep{srate2009praat} that automatically calculates the number of syllables per sample duration by detecting syllable nuclei \citep{wang2007robust} and pause duration. As for F0 extraction, we adopt an autocorrelation-based method \citep{Pitch-ac} implemented in Praat. We use gender-specific frequency ranges set to [75, 200] Hz for males and [100, 300] Hz for females. We initially tested F0 extraction with wider F0 ranges
but it was observed that the selected ranges were appropriate to exclude possible tracking errors and outliers in the F0 contour. The parameters to select the F0 candidates at 10ms intervals were set at their default values in Praat: silence threshold 0.03, voicing threshold 0.45, octave cost per octave 0.01, octave-jump cost 0.35, and voiced-unvoiced transition cost 0.14.

We summarize F0 values of each utterance using two summary statistics, \textcolor{\revcolor}{namely,} median and standard deviation. They reflect, respectively, the average pitch range and pitch dynamics within a given utterance. We study changes in these summary statistics between the zero-effort and mimicry attempts, with the aim of studying whether or not our attackers attempt to match their broad prosody characteristics with those of their targets upon their mimicry attempts.

\subsection{\textcolor{\revcolor}{Estimation and alignment of formant frequencies}}
\textcolor{\revcolor}{We performed formant analysis by comparing formant information of aligned utterances. First, we extracted formant center frequencies of the first three formants (F1, F2, and F3) using VoiceSauce \cite{Voicesauce} with Praat backend. Next, we aligned attacker's utterances (natural \& mimicry) with target's utterance using \emph{dynamic time warping} (DTW) \cite{sakoe1978dynamic}. The aligning process was done similarly as in \cite{vestman2018speaker}. This process involves using automatic selection of active speech frames that are well aligned and have reliable formant information. The alignment of utterances turned out to be challenging due to differences in speaking styles, acoustic conditions, and small deviations in spoken texts caused by mumbling. Thus, in addition to the automatic frame selection, we listened the aligned utterances in order to discard the the badly misaligned ones. Finally, after getting the aligned formant data, we measured the formant difference $d$ between utterances $a$ and $b$ as
\begin{equation}
\label{formant_diff_equation}
    d(a, b) = \frac{1}{3T} \sum_{t=1}^T \sum_{n=1}^3 \left| f_a(t, n) - f_b(t, n) \right|,
\end{equation}
where $T$ is the number of aligned frames and $f_a(t,n)$ is the center frequency of formant $n$ of utterance $a$ at frame $t$.}

\subsection{Results of prosody \textcolor{\revcolor}{and formant} analysis}

\begin{figure}[t]
\centering
\makebox[\linewidth][c]{
\begin{subfigure}{0.05\textwidth}
\includegraphics[height=9cm, trim={0 0 6.3cm 0cm}, clip]{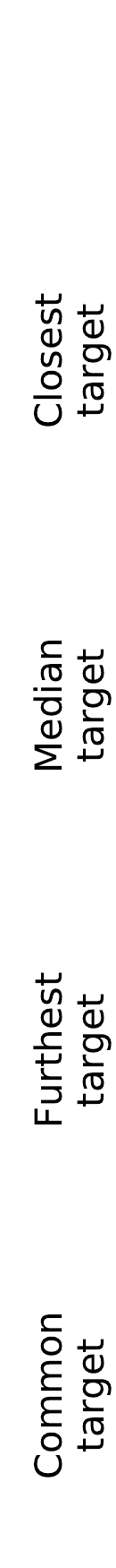}
\label{fig:dummy}
\end{subfigure}
\begin{subfigure}{0.3\textwidth}
\includegraphics[height=9cm]{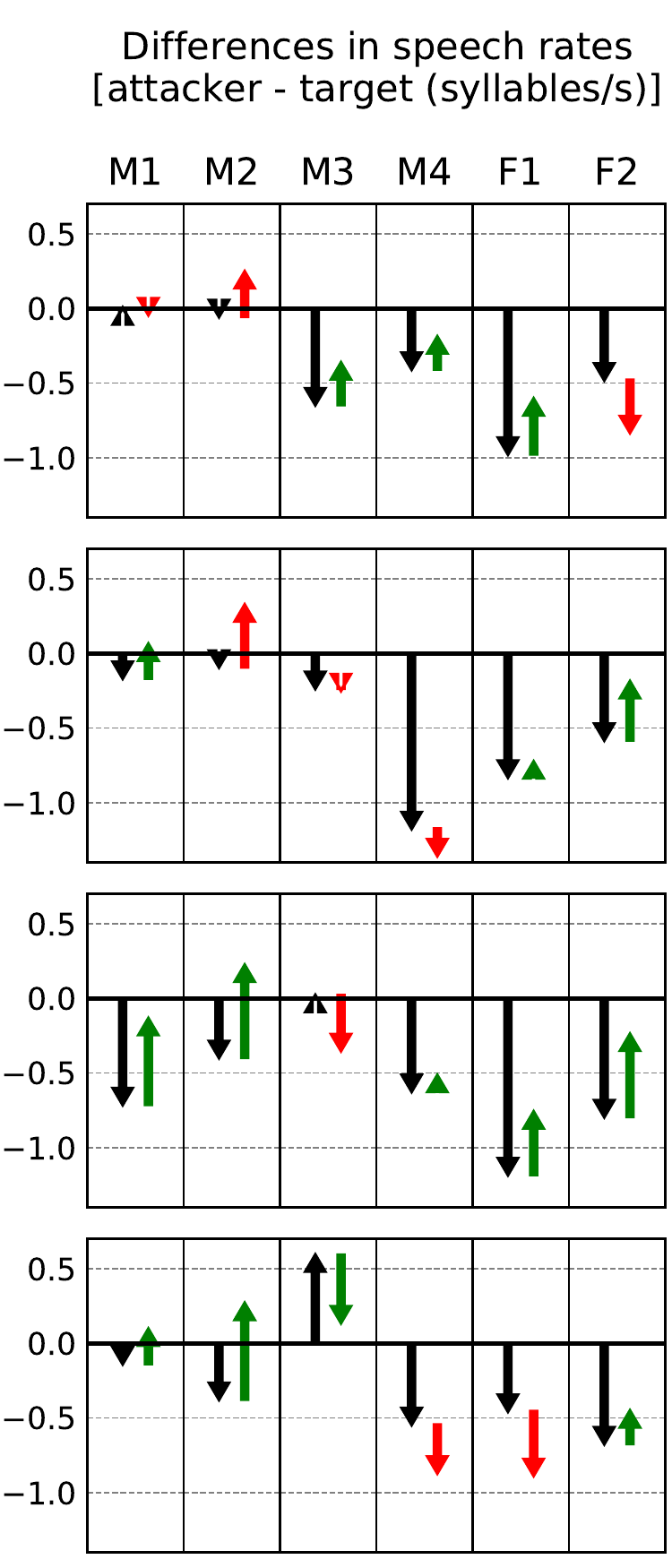}
\caption{}
\label{fig:speech_rate}
\end{subfigure}
\begin{subfigure}{0.3\textwidth}
\includegraphics[height=9cm]{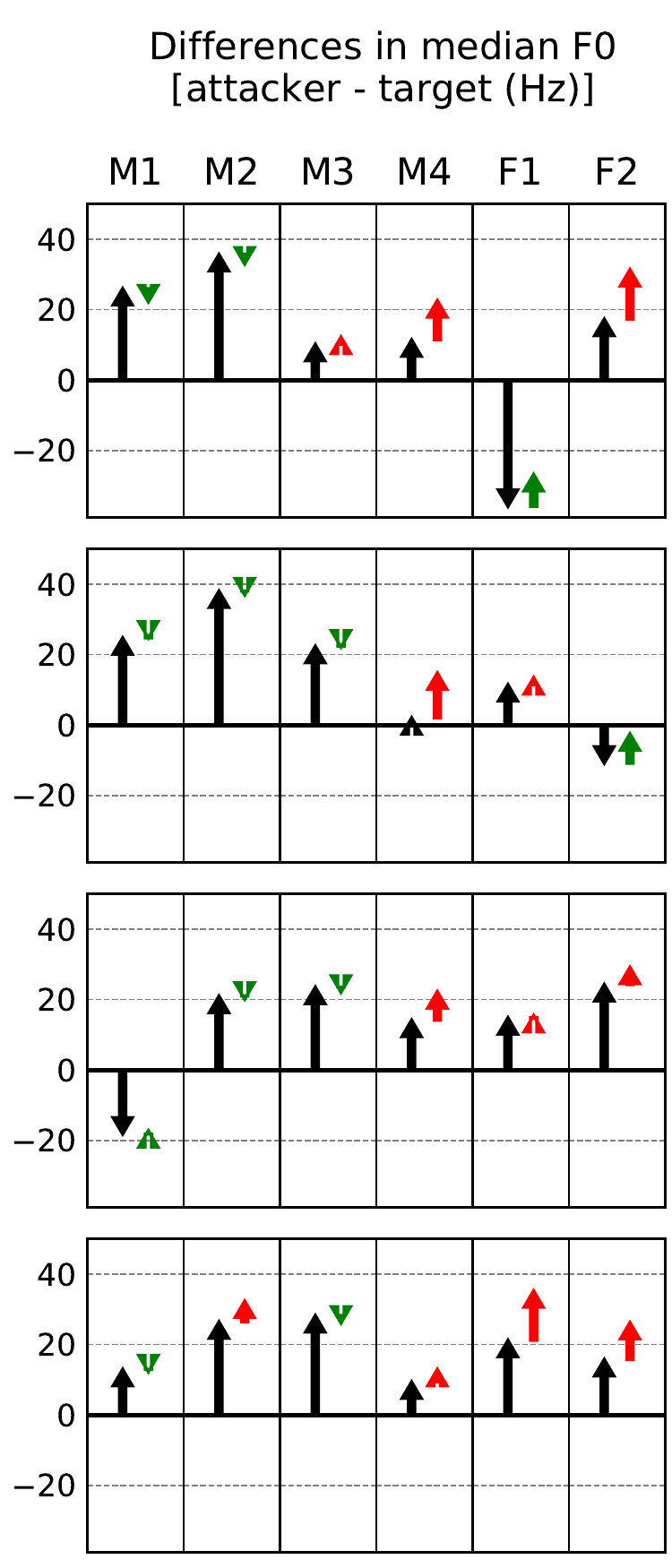}
\caption{}
\label{fig:f0}
\end{subfigure}
\begin{subfigure}{0.3\textwidth}
\includegraphics[height=9cm]{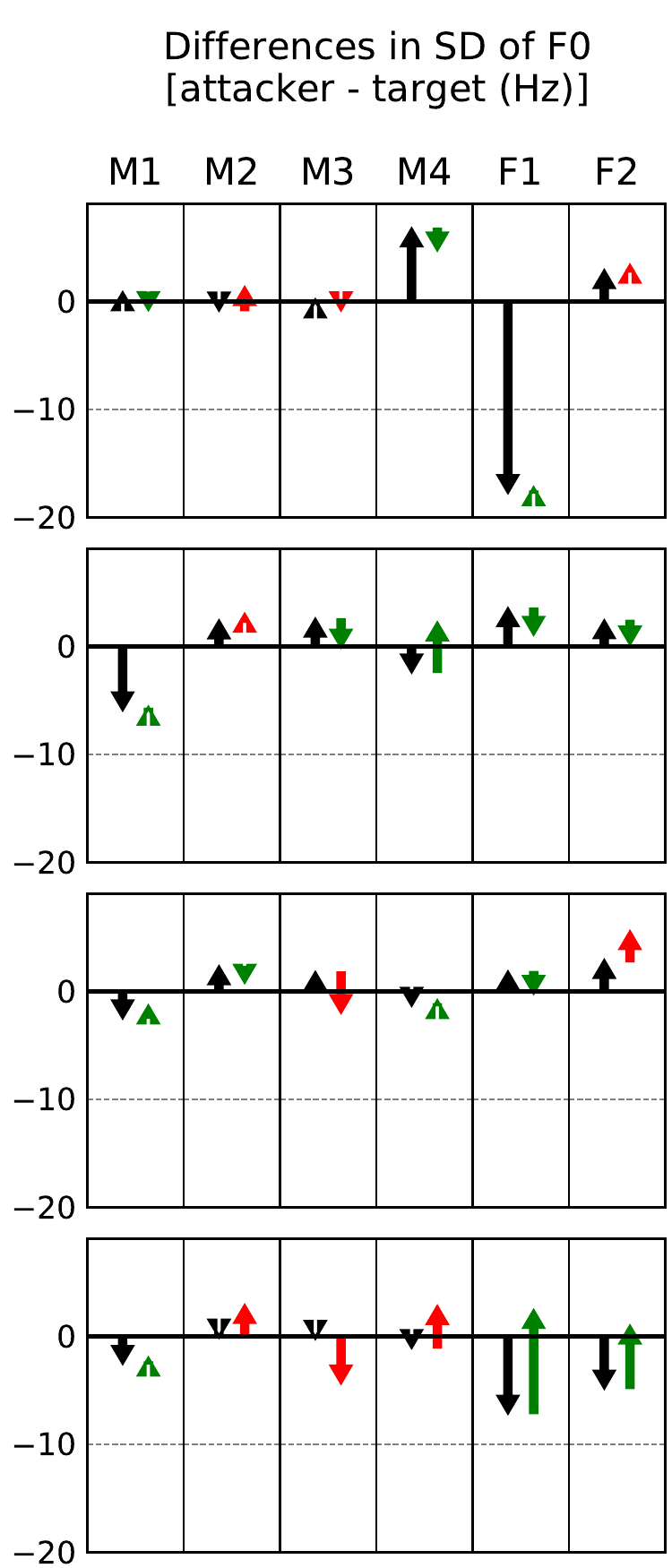} 
\caption{}
\label{fig:f0_sd}
\end{subfigure}
\begin{subfigure}{0.3\textwidth}
\includegraphics[height=9cm]{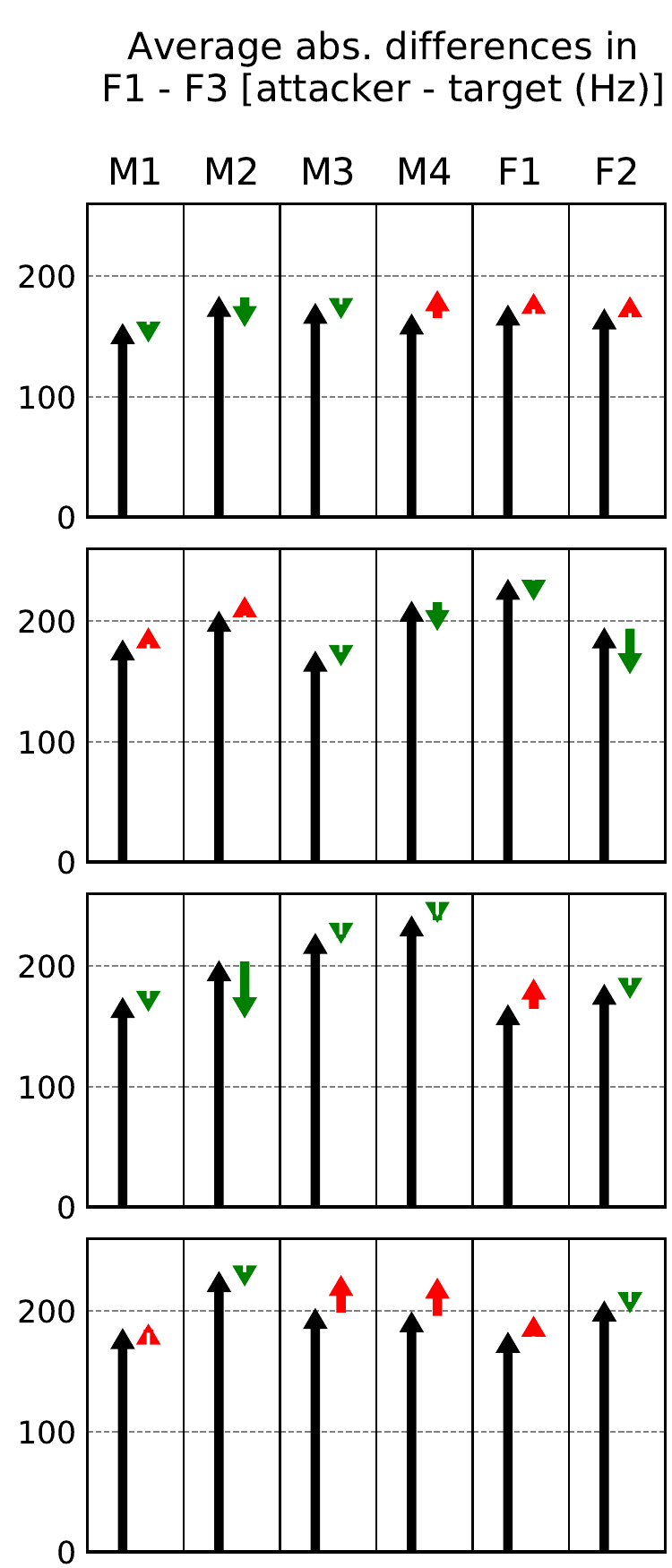} 
\caption{}
\label{fig:formants}
\end{subfigure}
}
\caption{Differences of attacker's (M1, M2, M3, M4, F1, F2) prosodic \textcolor{\revcolor}{and formant parameters to target's parameters} for all attacker-target combinations. Differences are shown for non-effort speech (\textbf{black arrow}) and for mimicked speech. The effect of mimicry is displayed with a \textbf{\textcolor{ForestGreen}{green arrow}} if it made attacker's and target's \textcolor{\revcolor}{parameters} closer to each other and with a \textbf{\textcolor{red}{red arrow}} otherwise.}

\end{figure}

In Figure \ref{fig:speech_rate}, we show the results for the analysis of speech rate differences. For each attacker-target combination, the displayed speech rates are obtained by averaging the speech rates of the available utterances (\textcolor{\revcolor}{4} to \textcolor{\revcolor}{7} utterances per combination). The results indicate that the speech rates of the attackers were, in general, slower than the targets' speech rates, when the attackers were not mimicking. This was anticipated, since the attackers were reading prompted text from a paper yielding slower speaking rates as opposed to those of the targets samples obtained from conversational situations. After listening to target's speech, the attackers were \textcolor{\revcolor}{in most cases} able to change their speech rates towards the targets' speech rates. At the individual level, we find that the male attacker \textcolor{\revcolor}{1} (M1) was good at \textcolor{\revcolor}{adjusting} his speech rate, while the male attacker 3 (M3) had naturally fast reading pace so that in some cases \textcolor{\revcolor}{(common target) his} speech rate \textcolor{\revcolor}{was already too fast}.

A similar comparison regarding F0 statistics is shown in Figures \ref{fig:f0} and \ref{fig:f0_sd}. We find that the attackers M1, M2, and M3 did not change their F0 considerably while mimicking, whereas attackers M4, F1, and F2 had some mimicry attempts with clearly different F0 than what their natural F0 is. \textcolor{\revcolor}{We do not observe clear differences between closest, median, and furthest target categories in terms of distances in F0 parameters between attackers and targets.}

\textcolor{\revcolor}{Finally, in Figure \ref{fig:formants}, we depict the formant differences between targets and attackers as defined in \eqref{formant_diff_equation}. Again, we find that the mimicking did not have major impact to the similarity of the formant frequencies. In 14 out of 24 cases, mimickers managed to get slightly closer to their targets in terms of the given metric. We further find that the formant differences are larger in the furthest category than in the closest category, which is expected as the location of formants affect the spectral features used in the target speaker selection.}

\section{Conclusion}

Biometric data uploaded to the Internet in large quantities, including human voice samples, opens up potential for misuse whenever the same biometric identifiers are adopted for strong user authentication to regulate access to personal data records, bank accounts and other services. Our study addressed a potential risk related to combination of public-domain automatic speaker verification (ASV) technology and public-domain voice \textcolor{\revcolor}{data}. The former is used as a search tool to identify potential target speakers to be mimicked.

Our results suggest that human mimicry is a rather special skill and less effective in spoofing modern ASV systems compared to voice conversion, text-to-speech, and replay. \textcolor{\revcolor}{In specific, none of our six attackers
received high detection scores for their attacks from our
simulated}\footnote{\textcolor{\revcolor}{The ASV implementations combine scripts/tools (\emph{e.g.} MSR Identity Toolkit, Kaldi) that are all public-domain code. They should be considered as proxies of modern ASV technology, rather than  end-user software.}} \textcolor{\revcolor}{public-domain or attacked ASV systems. Similar negative findings have been reported in earlier studies and are often speculated to be due to difficulty of humans to mimic accurately low-level spectral cues employed by ASV systems. One of our motivations was to \emph{re-assess} whether speech mimicry --- one of the weakest known attacks against ASV --- might be made substantially stronger (or more practical) when the target speakers are selected using ASV. We approached this question from two perspectives. On the one hand, we wanted to find out how the score ranges associated with broad target speaker rank (closest, median, further) transfer from the attacker's ASV to the attacked ASV. This is the \emph{technology} dimension of our attack model. On the other hand, we wanted to isolate the effect of the mimicry effort by collecting attackers' voice samples both `before' (zero-effort attack) and `after' (mimicry attack) listening to the target speaker's voice. This allows us to analyze the changes in attacker-to-target log-likelihood ratio (LLR) scores due to mimicry effect alone. This is the \emph{human} dimension of our attack model.} 
\textcolor{\revcolor}{Concerning the broad target speaker rank, the score relations generalize well from the attacker's ASV system to the attacked ASV system: LLR(closest target) $>$ LLR(median target) $>$ LLR(furthest target) relationship was retained both for Finnish and non-Finnish targets. This suggests that one could, indeed, use one ASV system (here, i-vector PLDA) to emulate the broad speaker ranking of another, targeted ASV system (here, x-vector PLDA). We find this result interesting and worthwhile of future work}. Even if the VoxCeleb corpora are among the largest (public) speaker corpora at this time, they are still tiny compared to the number of voice samples in the Internet. It would be interesting to repeat a similar study design to ours in a few years, perhaps with an order of magnitude larger target speaker corpus and, at this stage, unforeseen ASV technology. It would be important to uncover the conditions under which such emulation succeeds (or fails). With an increasing number of video and voice samples posted online, it is not only the security, but user privacy, that deserves attention.

\textcolor{\revcolor}{Concerning the impact of mimicry effort, the attacker-to-target LLRs remained low, and substantially below the target-to-target LLRs in both zero-effort and mimicry scenarios. Curiously, while the LLR scores for the furthest target speakers indicated some increase between zero-effort and mimicry scenarios, for the closest targets the LLR scores \emph{decreased} (but significantly only for the non-Finnish target speakers). To sum up, the broad target speaker rank generalized across the ASV systems, while the mimicry effect itself lead to negative (or no difference) effect. These findings reinforce the conjecture that voice mimicry by itself may not pose a strong attack against ASV; but ASV-based target speaker selection may.}

\textcolor{\revcolor}{We hypothesized that while our attackers' mimicry efforts did not have major impact on the ASV scores, they might have impact on human perception. Human listeners might, to some degree, focus on different cues of speaker identity than the ASV systems, which mostly focus on spectral characteristics of speech. However, the results of our listening test did not support the above hypothesis, as the results showed similar patterns to those we saw from the ASV scores.}

\textcolor{\revcolor}{So as to understand better the mimicry strategies implemented by the attackers, we also analyzed changes in formant frequencies and prosody statistics (F0, speaking rate). Even if some attackers were able to adjust their average formant frequencies towards those of their target speakers, the relative change in attacker-to-target formant distance (from zero-effort to mimicry) was minor. Adjustments in F0 statistics were minor as well. The most prominent adjustments towards the targets were seen in the speaking rate.}

\textcolor{\revcolor}{Our study has a number of limitations that one should take into account in future studies. First, the number of attackers (six) is admittedly small. This limitation, familiar to some of the authors \cite{Gonzalez2015-mimicry}, is common to most speech mimicry studies and relates to difficulties in data collection. The number of attackers varies from 1 to half dozen (or so) \cite{Wu2015-spoofing-survey}. Here, additional complications were caused by tailored target speaker selection, involving tedious speech transcription and several stages of data quality auditing. In future work, it might be practical to drop the transcription step and ask the attackers to impersonate their targets based on audio only. Another way to scale up the study would be attacker recruitment through crowdsourcing \cite{Panjwani2014-crowd}. This will, however, introduce new uncontrolled variations (such as attacker microphone differences). All our attacks were recorded using the same gear in the same room.}

\textcolor{\revcolor}{The second limitation relates to the  cross-domain data conditions: our attackers are native Finnish speakers, while VoxCeleb consists of many different nationalities and accents. Further, VoxCeleb consists of conversational speech while our attackers read text passages in an office environment. These differences induce style differences and might make the impersonation task harder for the attackers. This limitation is primarily due to lack of large Finnish celebrity corpus at the authors' exposure, as well as our preference to interact with the attackers conveniently. It would be interesting to repeat selected experiments using a larger target speaker corpus with matched mother tongue. In VoxCeleb, we are limited to 44 Finnish target speakers. Future work could therefore either adopt a larger Finnish celebrity corpus, or to recruit native American English attackers. Given the nature of  \emph{found data}, controlling all the variations will be difficult.}

Our attacks could also be made stronger in a number of ways.
First, the attacker might use the public-domain ASV system in a more proactive way, such as optimizing its detection accuracy further in off-line experiments. Second, the attacker could potentially utilize \textcolor{\revcolor}{more detailed feedback from a dedicated ASV system --- in this work, attackers used ASV for speaker \emph{ranking} while some prior work has used ASV score
as a feedback signal \cite{Zetterholm2004-PerceptionSpeakerVerification}}.
Third, assuming there would be an actual monetary (or other strong) motivator to seriously mimic someone --- similar to practicing to forge someone's signature --- the attacker might use substantially more effort to get familiar with the speaking style of his or her targets. He or she might perhaps use feedback from prosody measurements in addition to ASV score. In our study, given the extensive work required to prepare the tailored targets and collect the data, all the above had to be relaxed to complete recordings in a reasonable time. The mimicry attacks (with audio reference of the target) took place in a single session and our attackers completed their mimicry tasks relatively fast. Nonetheless, in future work it would be interesting to evaluate whether mimicry attacks could be improved with further, and more proactive, training. Another interesting target would be studying combination of automatic target speaker selection with voice conversion (or other technical) spoofing attacks.

\textcolor{\revcolor}{It would be also interesting to address whether, and how, one may benefit from current (or suitably modified) ASV methods to provide intuitive feedback to improve one's mimicry skills. This would be potentially helpful in suggesting specific articulatory or voice source modifications required to increase the ASV score. The present study was framed to the context of ASV attacks but such methods could be potentially useful for mimicry artists, voice actors, and language learners as well.}

\section*{Acknowledgement}
The work has been supported by Academy of Finland (proj. no. 309629 entitled ``NOTCH: NOn-cooperaTive speaker CHaracterization'') and by the Doctoral Programme in Science, Technology and Computing (SCITECO) of the UEF. \textcolor{\revcolor}{A part of the work of the first author was supported by NEC internship program.} The work of Md Sahidullah was made with the support of Region Grand Est. We gratefully acknowledge the support of NVIDIA Corporation with the donation of the Titan V GPU used for this research. 

\section*{References}

\bibliographystyle{elsarticle-num}
\bibliography{refs}

\end{document}